\documentclass[showpacs,
floatfix,
aps,
prl,
amsmath,
twocolumn,
superscriptaddress,
tightenlines]{revtex4-1}
\usepackage{graphicx}
\usepackage{dcolumn}
\usepackage{amsmath}
\usepackage{amsfonts}
\usepackage{bm}
\usepackage{color}
\usepackage{times}
\newcommand{\be}{\begin{equation}}
\newcommand{\ee}{\end{equation}}
\newcommand{\bea}{\begin{eqnarray}}
\newcommand{\eea}{\end{eqnarray}}

\def\lb{\left [}
\def\rb{\right ]}
\def\pc{\mathcal{P}}

\def\eac{\epsilon_{\omega}}
\def\edc{\epsilon_{j}}

\def\mac{m^\star}
\def\mdc{m^\star}
\def\fdc{f_j}

\def\oc{\omega_{\mbox{\scriptsize {c}}}}

\def\rc{R_{\mbox{\scriptsize {c}}}}

\def\tq{\tau_{\mbox{\scriptsize {q}}}}

\newcommand{\req}[1]{Eq.\,(\ref{#1})}

\newcommand{\rfig}[1]{Fig.\,\ref{#1}}
\newcommand{\rFig}[1]{Figure \ref{#1}}
\newcommand{\rref}[1]{Ref.\,\onlinecite{#1}}

\usepackage[english,american]{babel}
\begin{document}

\title{Hall field-induced resistance oscillations in a $p$-type Ge/SiGe quantum well}

\author{Q.~Shi}
\affiliation{School of Physics and Astronomy, University of Minnesota, Minneapolis, Minnesota 55455, USA}
\author{Q.~A.~Ebner}
\affiliation{School of Physics and Astronomy, University of Minnesota, Minneapolis, Minnesota 55455, USA}
\author{M.~A.~Zudov}
\affiliation{School of Physics and Astronomy, University of Minnesota, Minneapolis, Minnesota 55455, USA}
\author{O.~A.~Mironov}
\affiliation{Department of Physics, University of Warwick, Coventry, CV4 7AL, United Kingdom}
\affiliation{International Laboratory of High Magnetic Fields and Low Temperatures, 53-421 Wroclaw, Poland}
\author{D.~R.~Leadley}
\affiliation{Department of Physics, University of Warwick, Coventry, CV4 7AL, United Kingdom}
\begin{abstract}
We report on a magnetotransport study in  a high-mobility 2D hole gas hosted in a pure Ge/SiGe quantum well subject to dc electric fields and high frequency microwave radiation. 
We find that under applied dc bias the differential resistivity exhibits a pronounced maximum at a magnetic field which increases linearly with the applied current.
We associate this maximum with the fundamental peak of Hall field-induced resistance oscillations (HIRO) which are known to occur in 2D electron gases in GaAs/AlGaAs systems.
After taking into account the Dingle factor correction, we find that the position of the HIRO peak is well described by the hole effective mass $\mdc \approx 0.09\,m_0$, obtained from microwave photoresistance in the same sample.
\end{abstract}

\received{April 16, 2014}
\maketitle

Magnetotransport in very high Landau levels of high-mobility two-dimensional electron gases (2DEG) hosted in GaAs/AlGaAs quantum wells is known to exhibit a variety of fascinating phenomena \citep{dmitriev:2012}.
For example, when a 2DEG is exposed to microwave radiation and weak perpendicular magnetic field $B$, the magnetoresistance acquires prominent oscillations \citep{zudov:2001a,ye:2001}, which are controlled by $\eac = \omega/\oc$, where $\omega = 2\pi f$ is the microwave frequency and $\oc=eB/m^\star$ is the cyclotron frequency of the charge carrier with the effective mass $m^\star$. 
These oscillations, known as microwave-induced resistance oscillations (MIRO), are usually explained in terms of the displacement mechanism \citep{ryzhii:1970,durst:2003,lei:2003,vavilov:2004}, which originates from the radiation-induced modification of scattering off impurities, and the inelastic mechanism \citep{dorozhkin:2003,dmitriev:2005}, which stems from the radiation-induced changes in the distribution function.
Both mechanisms predict that the photoresistivity $\delta\rho_\omega$ oscillates as 
\be
\delta \rho_\omega/\rho_D = - 2\pi \eta \pc \lambda^2 \eac \sin 2\pi\eac\,,
\label{eq.miro}
\ee
where $\rho_D$ is the Drude resistivity, $\eta$ is the dimensionless scattering rate, which contains both displacement and inelastic contributions \citep{dmitriev:2009b}, $\pc$ is the dimensionless microwave power \citep{dmitriev:2005,khodas:2008}, $\lambda = \exp(-\pi/\oc\tq)$ is the Dingle factor, and $\tq$ is the quantum lifetime.
In extremely clean 2DEG \citep{note:1}, the MIRO minima can develop into zero-resistance \citep{mani:2002,zudov:2003} or zero-conductance \citep{yang:2003} states.
Microwave-induced magneto-conductance oscillations and associated zero-conductance states \citep{yang:2003} have been also realized in a non-degenerate 2D system, electrons on liquid helium surface \citep{konstantinov:2009,konstantinov:2010}.
Most recently, MIRO have been observed in a two-dimensional hole gas (2DHG) in Ge/SiGe quantum wells \citep{zudov:2014}.

Another class of magneto-oscillations appears in the differential resistivity $r$ of a Hall bar-shaped structures under applied direct current \citep{yang:2002,bykov:2005c,zhang:2007a}.
These oscillations, known as Hall field-induced resistance oscillations (HIRO), originate exclusively from the displacement mechanism \citep{vavilov:2007,lei:2007} as a result of the commensurability between the cyclotron diameter $2\rc$ and the spatial separation between Laudau levels,
tilted in space by Hall electric field $E = \rho_H j$, where $j$ is the current density and $\rho_H$ is the Hall resistivity.
More specifically, whenever $\edc = eE(2\rc)/\hbar\oc$ is close to an integer value, the probability for an electron to make an elastic transition to a higher Landau level as a result of backscattering off short range disorder is maximized, giving rise to a maximum in $r$.
At $2\pi\edc \gg 1$, the resultant correction to the differential resistivity $\delta r$ is given by
\be
\frac {\delta r}{\rho_D} =\frac{16}{\pi}\frac{\tau}{\tau_{\pi}}\lambda^2 \cos 2\pi \edc \,,
\label{eq.hiro}
\ee
where $\tau$ is the transport scattering time and $\tau_{\pi}$ is the backscattering time.
As suggested by \req{eq.hiro}, HIRO can be used to obtain the effective mass of the charge carrier.
This can be achieved using, e.g., the dependence of the magnetic field of the fundamental HIRO peak ($\edc = 1$) on $j$, 
\be
B_1 \approx \frac{4\pi m^\star}{e^2k_F} j\,,
\label{b1}
\ee
where $k_F$ is the Fermi wavenumber.
To date, studies of HIRO have been limited almost exclusively \citep{note:2} to high mobility 2DEG in GaAs/AlGaAs heterostructures \citep{yang:2002,bykov:2005c,zhang:2007a,bykov:2007,zhang:2008,hatke:2009c,hatke:2010a,hatke:2011a}.
While \req{b1} was successfully employed in the above studies, it disregards the $B$-dependence of $\lambda^2$ which, as we show below, becomes important at sufficiently low $\tq$ values.

In this Rapid Communication we report on a nonlinear magnetotransport study in a new material system, a high-mobility two-dimensional hole gas hosted in a pure Ge/SiGe quantum well \citep{dobbie:2012,hassan:2013}. 
Under applied dc bias, the differential resistivity exhibits a pronounced maximum which shifts to higher $B$ with increasing $j$.
In agreement with \req{b1}, we observe a roughly linear relationship between the peak position $B_1$ and $j$.
However, direct employment of \req{b1} yields an estimate for the hole effective mass of $\mdc \approx 0.11\,m_0$, noticeably larger than $\mac \approx 0.09\,m_0$ obtained in recent MIRO experiments \citep{zudov:2014}. 
To investigate this discrepancy, we have measured microwave photoresistance in the same sample, from which we have obtained $\mac \approx 0.087 \,m_0$, in good agreement with \rref{zudov:2014}.
High quality of our MIRO data also allowed us, for the first time, to perform a Dingle analysis from which we obtained $\tq \approx 2.8$ ps.
We demonstrate that in our 2DHG, \req{b1} underestimates the position of the fundamental HIRO maximum due to (i) the strong $B$-dependence of the Dingle factor, which is neglected in \req{b1}, and (ii) the approximate nature of \req{eq.hiro} near $\edc \approx 1$.
Once the above factors are taken into account, we find that HIRO in our 2DHG are well described by $\mdc \approx 0.09\,m_0$.

Our lithographically defined 50 $\mu$m-wide Hall bar sample was fabricated from a fully strained, 20 nm-wide, Ge/Si$_{0.2}$Ge$_{0.8}$ quantum well grown by reduced pressure chemical vapor deposition \citep{dobbie:2012}. 
Holes were supplied by a 10 nm-wide Boron-doped layer separated from the Ge channel by a 26 nm-wide undoped Si$_{0.2}$Ge$_{0.8}$ spacer. 
At $T \approx 1$ K, our 2DHG has the hole density $p \approx 2.8 \times 10^{11}$ cm$^{-2}$ and the mobility $\mu \approx 1.3 \times 10^6$ cm$^2$/Vs. 
The resistivity $\rho$ and the differential resistivity $r$ were measured in sweeping $B$ using a standard four-terminal lock-in technique.

\begin{figure}[t]
\includegraphics{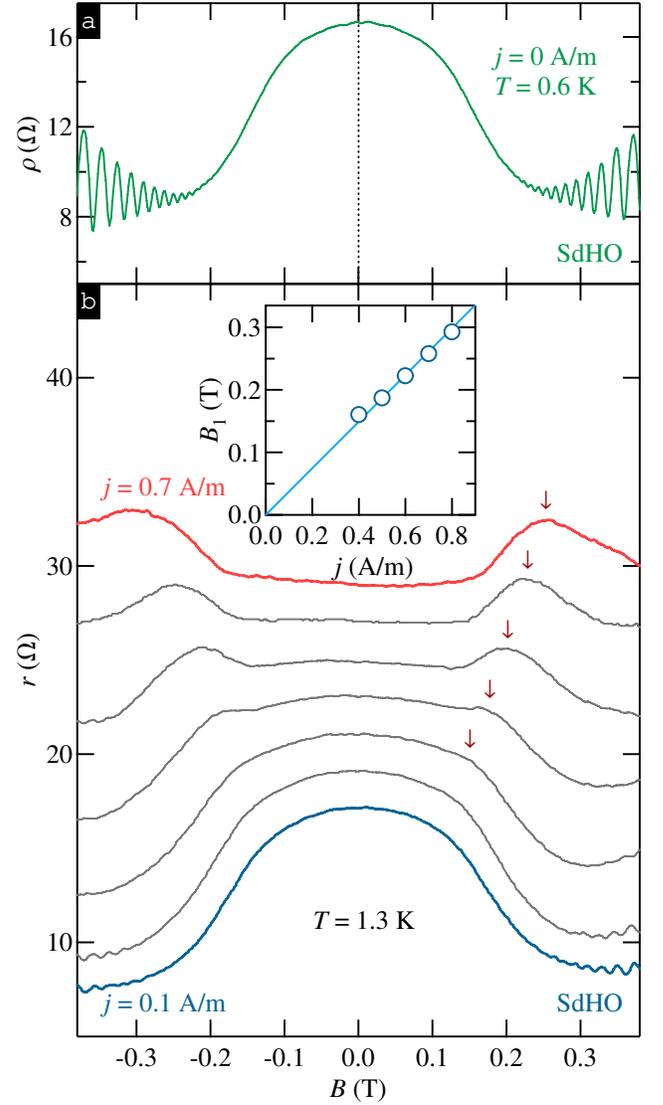}
\vspace{-0.05 in}
\caption{(Color online)
(a) Magnetoresistivity $\rho(B)$ measured at $T \approx 0.6$ K.
(b) Differential resistivity $r(B)$ at current densities from $j$ = 0.1 A/m to 0.7 A/m, in steps of 0.1 A/m, measured at $T \approx 1.3$ K.
The traces are vertically offset for clarity by 2 $\Omega$.
The inset shows the position of the fundamental HIRO peak $B_1$ as a function of the current density $j$.
Fit to $B_1 = 4\pi \mdc j/e^2k_F$ yields $\mdc \approx 0.11 \,m_0$.
}
\vspace{-0.15 in}
\label{hiro}
\end{figure}

In \rfig{hiro}(a) we present magnetoresistivity $\rho (B)$, measured at $T \approx 0.6$ K, which exhibits significant negative magnetoresistance effect, similar to what have been recently observed in 2DEG in GaAs/AlGaAs quantum wells \citep{dai:2010,dai:2011,hatke:2011b,bockhorn:2011,hatke:2012a,bockhorn:2014,shi:2014a}, and Shubnikov-de Hass oscillations.
\rFig{hiro}(b) shows the differential resistivity $r$ as a function of $B$ under applied direct currents with densities from $j = 0.1$ A/m to 0.7 A/m, in steps of 0.1 A/m, measured at $T \approx 1.3$ K.
Distinct peaks (cf. $\downarrow$) start showing up at $j$ = 0.3 A/m, symmetrically at both magnetic field directions.
As expected for HIRO, the peaks move to higher $B$ and grow in magnitude with increasing current.

According to \req{eq.hiro} the fundamental HIRO maximum should occur close to $\edc = 1$ and, as a result, its position should scale linearly with $j$, see \req{b1}. 
In the inset of \rfig{hiro}(b) we plot $B_1$ as a function of $j$ and observe the expected linear dependence.
The fit to the data using \req{b1} yields an effective mass of $\mdc \approx 0.11 \,m_0$.
This value is about 20\% higher than the effective mass, $\mac \approx 0.09$, obtained in a recent MIRO experiment \citep{zudov:2014} on a lower mobility Ge/SiGe quantum well.

There are at least two factors which might lead to an overestimated value of $\mdc$ obtained from our HIRO data using \req{b1}.
First, \req{eq.hiro} is valid only in the limit of $2\pi\edc \gg 1$, a condition which is only marginally met at $\edc \approx 1$.
For a more accurate description of HIRO, $\cos2\pi\edc$ in \req{eq.hiro} should be replaced by \citep{vavilov:2007}
\be
%\mathcal{F}(\edc) \approx \frac{2}{\pi}  \lb J_{1}^2(\pi\edc) - \lp 2\pi\edc- \frac 1 {2\pi\edc} \rp J_{0}(\pi\edc)J_{1}(\pi\edc) \rb\,.
\mathcal{F}_j(\edc) \approx \frac{\pi}{2}  \lb J_{1}^2(\pi\edc) - 2\pi\edc J_{0}(\pi\edc)J_{1}(\pi\edc) \rb\,,
\label{eq.f}
\ee
where $J_0$ and $J_1$ are the Bessel functions of the first kind.
The second factor which contributes to an overestimated value of $\mdc$ stems from the dependence of the Dingle factor on $\edc$, $\lambda^2(\edc) = \exp (-\edc/\fdc\tq)$, where $\fdc = 2 j/e k_F$.
%%%%%%%%%%%%%%%%%%%%%%%%%%%%%%%%%%%%%%%%%%%%
\begin{figure}[t]
\includegraphics{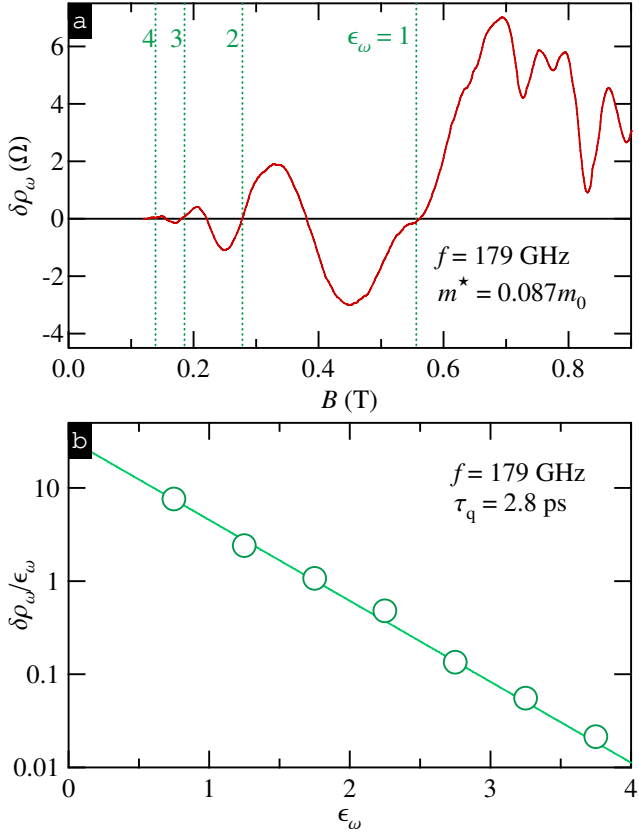}
\vspace{-0.05 in}
\caption{(Color online) 
(a) Microwave photoresistance $\delta \rho_\omega$ as a function of $B$ measured at $f = 179$ GHz and $T = 0.6$ K.
Vertical lines are drawn at the cyclotron resonance harmonics, $\eac = 1,2,3,4$ calculated using $\mac = 0.087 \,m_0$.
(b) Reduced MIRO amplitude $\delta\rho_\omega/\eac$ as a function of $\eac$ on a log-linear scale.
The fit to $\exp(-\eac/f\tq)$ generates $\tq \approx 2.8$ ps.
}
\vspace{-0.15 in}
\label{miro}
\end{figure}
%%%%%%%%%%%%%%%%%%%%%%%%%%%%%%%%%

In order to more accurately access the value of $\edc$ which corresponds to the position of the fundamental HIRO peak $B_1$, it is necessary to know the quantum lifetime $\tq$.
In contrast to the HIRO experiments on 2DEG in high mobility GaAs/AlGaAs quantum wells, which routinely exhibit multiple oscillations \citep{zhang:2007a,zhang:2007c,hatke:2009c,hatke:2011a}, the data in our 2DHG reveal only one HIRO maximum and are not of sufficient quality to perform systematic Dingle analysis.
We can, however, obtain $\tq$ from microwave photoresistance, as discussed below.

In \rfig{miro}(a) we present microwave-induced resistance oscillations \citep{note:7} measured at $f = 179$ GHz and $T \approx 0.6$ K in the same sample.
We note that in our previous MIRO study \citep{zudov:2014}, which used a lower mobility ($\mu \approx 0.4 \times 10^6$ cm$^2$/Vs) device and lower microwave frequencies ($f \le 110$ GHz), only two MIRO maxima and one minimum were resolved.
In this experiment, however, we detect four pairs of maxima and minima, occurring on the opposite sides of the corresponding cyclotron resonance harmonics.
According to \req{eq.miro}, the photoresistance is expected to vanish at integer values of $\eac$, providing an accurate way to obtain the effective mass value.
As shown in \rfig{miro}(a), vertical lines which are drawn at $\eac = 1, 2,3$, and 4, calculated using $\mac = 0.087 \,m_0$ cross all observed zero-response nodes, $\delta\rho_\omega =0$.
The obtained value is in good agreement with an earlier estimate of $\mac = 0.09 \,m_0$ obtained in \rref{zudov:2014}.

The high quality of the photoresistance data shown in \rfig{miro}(a) allows us, for the first time, to perform a proper Dingle analysis of the MIRO amplitude in 2DHG hosted in Ge/SiGe quantum well.
In \rfig{miro}(b) we present a reduced MIRO amplitude $\delta\rho_\omega/\eac$ as a function of $\eac$ using a log-linear scale and observe well-behaved exponential dependence over more than two orders of magnitude.
The fit to the data using $\exp(-\eac/f\tq) \equiv \exp(-2\pi/\oc\tq)$ yields $\tq \approx 2.8$ ps.
This value is considerably lower than $\tq$ in high-mobility 2DEG in GaAs/AlGaAs, where it ranges between 10 and 20 ps \citep{zhang:2007a,hatke:2008a,hatke:2011b,hatke:2011f,hatke:2012d}.
Using $m^\star = 0.09 \,m_0$ and $\tq = 2.8$ ps we can estimate $\lambda^2$ at $B = 0.13$ T, the field where one can expect to observe the second HIRO maximum at current density $j = 0.7$ A/m \citep{note:4}. 
The obtained value of $\lambda^2 \sim 10^{-4}$ is about two orders of magnitude smaller than $\lambda^2$ at $B = 0.25$ T, explaining why no second HIRO maximum is detected.
This observation is consistent with the fact that MIRO also cease to exist at $B \lesssim 0.13$ kG (see also \rref{zudov:2014}).

%%%%%%%%%%%%%%%%%%%%%%%%%%%%%%%%%%%%%%%%%%%%
\begin{figure}[t]
\includegraphics{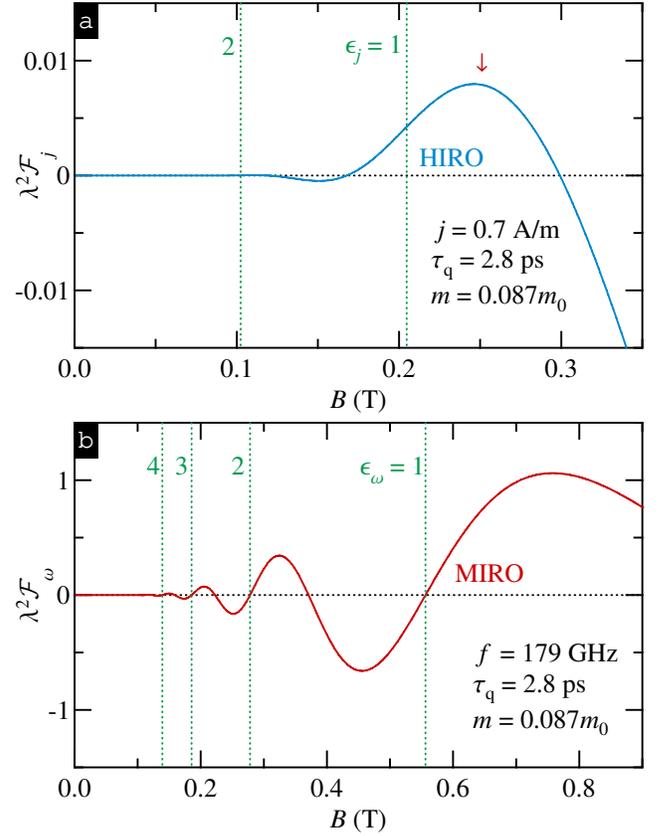}
\vspace{-0.05 in}
\caption{(Color online) 
(a) $\lambda^2\mathcal{F}_j$ calculated using \req{eq.f} with $j = 0.7$ A/m, $\tq = 2.8$ ps, $\mdc = 0.087\,m_0$ as a function of $B$.
Vertical lines are drawn at $\edc =1, 2$.
Notice that the fundamental HIRO peak (cf. $\downarrow$) occurs at $B$ which is 20\,\% higher than $B_1$ given by \req{b1} (cf. vertical line marked by $\edc = 1$).
(b) $\lambda^2\mathcal{F}_\omega = -2\pi\eac\lambda^2\sin 2\pi\eac$ calculated with $f = 179$ GHz, $\tq = 2.8$ ps, $\mdc = 0.087\,m_0$ as a function of $B$.
Vertical lines are drawn at $\eac = 1,2,3,4$.
}
\vspace{-0.15 in}
\label{calc}
\end{figure}
%%%%%%%%%%%%%%%%%%%%%%%%%%%%%%%%%
Having obtained the quantum lifetime, we now demonstrate that incorporating the $B$-dependence of the Dingle factor and using \req{eq.f}, instead of approximate expression given by \req{eq.hiro}, can indeed explain larger value of the effective mass obtained using \req{b1}.

In \rfig{calc}(a) we present $\lambda^2\mathcal{F}_j$ calculated using \req{eq.f} with $j = 0.7$ A/m, $\tq = 2.8$ ps, $\mdc = 0.087\,m_0$ as a function of $B$.
The vertical lines are drawn at $\edc =1, 2$.
As one can see, only a single HIRO peak is observed, in agreement with our experimental data.
Furthermore, this peak occurs at $B \approx 0.25$ T, which is very close to what we observe in experiment [cf. the top trace in \rfig{hiro}(b)] and is almost exactly 20\,\% higher than $B_1$ obtained from \req{b1}.
Based on this observation, we conclude that both MIRO and HIRO are well described by $\mdc \approx 0.09\,m_0$.
We emphasize that to obtain accurate position of the fundamental HIRO peak it is important to both include the Dingle factor correction and to use \req{eq.f}, instead of approximate \req{eq.hiro}. 
We also notice that the Dingle factor correction does not affect the accuracy of $\mac$ obtained from MIRO since we are using zero-response nodes which, according to \req{eq.miro} exactly correspond to integer values of $\eac$.

Finally, to qualitatively confirm the value of $\tq$, obtained from the Dingle plot in \rfig{miro}(b), we present in \rfig{calc}(b) $\lambda^2\mathcal{F}_\omega = -2\pi\eac\lambda^2\sin 2\pi\eac$ calculated using $f = 179$ GHz$, \tq = 2.8$ ps, and $\mdc = 0.087\,m_0$ as a function of $B$.
We observe that the calculated curve very well reproduces our experimental findings; same number of oscillations, same magnetic field onset, and same locations of zero-response nodes (cf. vertical lines).

%summary
In summary, we have studied nonlinear magnetotransport in a high-mobility two-dimensional hole gas hosted in a pure Ge/SiGe quantum well. 
Under applied dc bias, the differential resistivity shows a pronounced maximum which moves to higher magnetic fields with increasing direct current.
We associate this maximum with the fundamental peak of Hall field-induced resistance oscillations (HIRO), which are frequently observed in 2DEG in GaAs/AlGaAs heterostructures.
Our analysis shows that the position of the HIRO peak and microwave-induced resistance oscillations observed in the same sample are both well described by the hole effective mass $\mdc \approx 0.09\,m_0$.
Since previous studies of MIRO and HIRO were limited exclusively to GaAs/AlGaAs systems, our findings establish that a strained Ge/SiGe quantum well is a promising 2D hole system, with a light carrier mass and high mobility, for further exploration of nonequilibrium transport phenomena.

%thanks
We thank I. A. Dmitriev for discussions, A. Dobbie for Ge/Si$_{0.2}$Ge$_{0.8}$ wafer growth, A. H. A. Hassan for device fabrication, R. J. H. Morris for SIMS/XRD characterization and G. Jones, S. Hannas, T. Murphy, J. Park, and D. Smirnov for technical assistance with experiments.
The work at University of Minnesota was funded by the NSF Grant No. DMR-1309578 (measurements of differential resistivity) and by the  US Department of Energy, Office of Basic Energy Sciences, under Grant No. ER 46640 – SC0002567 (measurements of microwave photoresistance). 
The work at Warwick University was partially supported by EPSRC projects EP/F031408/1 and EP/J001074/1. 
O.A.M. acknowledges the support of ILHMFLT (Wroclaw, Poland), SAS Centre of Excellence CFNT MVEP (Kosice, Slovakia), DFG project DR832/3-1 at HLD-HZDR (Rossendorf, Germany) as a member of the European Magnetic Field Laboratory (EMFL), and National Scholarship Program of the Slovak Republic for the Mobility Support for Researchers in the Academic Year 2012/2013.
A portion of this work was performed at the National High Magnetic Field Laboratory, which is supported by NSF Cooperative Agreement No. DMR-0654118, by the State of Florida, and by the DOE and at the Center for Integrated Nanotechnologies, a U.S. Department of Energy, Office of Basic Energy Sciences user facility.  

%\bibliography{../../bibRMP1,footnotes_nmr}
%\bibliographystyle{../../apsrev}
%\bibliography{../../bibRMP1,footnotes}

\begin{thebibliography}{47}
\bibitem[{not({\natexlab{c}})}]{note:7}
\bibinfo{note}{Photoresistance was obtained by removing a smooth background
  from $\rho(B)$ measured under microwave irradiation.}
\bibitem[{not({\natexlab{d}})}]{note:4}
\bibinfo{note}{At higher currents, contacts to our 2DHG become non-Ohmic
  precluding reliable observation of HIRO.}
\end{thebibliography}

\end{document}